\documentclass[a4paper]{article}

\usepackage[backend=biber, style=nature, sorting=none, url=false, isbn=false, eprint=false, natbib=false, hyperref=true]{biblatex}

\usepackage{pdfpages,multirow,ragged2e}
\usepackage{comment}
\usepackage{amsmath}
\usepackage{amssymb}
\usepackage{hyperref}
\usepackage[USenglish]{babel}
\usepackage{authblk}
\usepackage[utf8]{inputenc}    
\usepackage{csquotes}

\newcommand{\fgt}{Fe$_{5}$GeTe$_2$ }
\newcommand{\fgtnospace}{Fe$_{5}$GeTe$_2$}

\let\oldref\ref\renewcommand{\ref}[1]{(\oldref{#1})} 

\makeatletter
	\ifboolexpr{bool{xetex}}
	 {\renewcommand{\Gin@extensions}{.pdf, .png,.jpg,.bmp,.pict,.tif,.psd,.mac,.sga,.tga,.gif, .eps,.ps,}}{}
\makeatother

\ifboolexpr{bool{xetex} or bool{luatex}}
 {}

\begin{document}

\title{Skyrmionic Spin Structures in Layered \fgt Up To Room Temperature}

\author[1]{Maurice Schmitt}
\author[2]{Thibaud Denneulin}
\author[2]{András Kovács}
\author[1]{Tom G. Saunderson}
\author[3, 4]{Philipp Rüßmann}
\author[1]{Aga Shahee}
\author[5]{Tanja Scholz}
\author[2]{Amir Tavabi}
\author[6]{Martin Gradhand}
\author[7]{Phivos Mavropoulos}
\author[5]{Bettina Lotsch}
\author[2]{Rafal Dunin-Borkowski}
\author[4]{Yuriy Mokrousov}
\author[4]{Stefan Blügel}
\author[1, 8]{Mathias Kläui}

\affil[1]{Johannes Gutenberg Universität Mainz, Institut für Physik, Staudingerweg 7, 55128 Mainz, Germany}
\affil[2]{Ernst Ruska Centre for Microscopy and Spectroscopy with Electrons and Peter Grünberg Institute, Forschungszentrum Jülich, 52425 Jülich, Germany}
\affil[3]{Institute of Theoretical Physics and Astrophysics,
University of Würzburg, Am Hubland, 97074 Würzburg, Germany}
\affil[4]{Peter Grünberg Institut and Institute for Advanced Simulation, Forschungszentrum Jülich, 52425 Jülich, Germany}
\affil[5]{Max Planck Institute for Solid State Research, Heisenbergstraße 1, 70569 Stuttgart, Germany}
\affil[6]{University of Bristol, School of Physics, HH Wills Physics Laboratory, Tyndall Avenue, Bristol BS8 1TL, England}
\affil[7]{National and Kapodistrian University of Athens, Department of Physics, University Campus, GR-157 84 Zografou, Athens}
\affil[8]{QuSpin, Norwegian University of Science and Technology, Department of Physics, NTNU NO-7491, Trondheim, Norway }
	
\maketitle
\begin{abstract}

The role of the crystal lattice, temperature and magnetic field for the spin structure formation in the 2D van der Waals magnet \fgt is a key open question. Using Lorentz transmission electron microscopy, we experimentally observe topological spin structures up to room temperature in the metastable pre-cooling and stable post-cooling phase of \fgtnospace. Over wide temperature and field ranges, skyrmionic magnetic bubbles form without preferred chirality, which is indicative of a centrosymmetric crystal structure. In the pre-cooling phase, these bubbles are observable even without the application of an external field, while in the post-cooling phase, a transformation from bubble domains to stripe domains is seen. To understand the magnetic order in \fgt we compare macroscopic magnetometry characterization results with microscopic density functional theory calculation. Our results show that even up to room temperature, topological spin structures can be stabilized in centrosymmetric van der Waals magnets.


\end{abstract}

\section{Introduction}
Two-dimensional (2D) magnetic van der Waals (vdW) materials have seen rising attention since their recent introduction into the field of spintronics and beyond \cite{Gibertini.2019}. Many of these materials show promising properties for future applications, such as the semiconducting 2D ferromagnet CrI$_{3}$ \cite{Huang.2017}, the metallic 2D ferromagnet Fe$_{3}$GeTe$_2$ \cite{Deng.2018} and the very recently discovered metallic 2D ferromagnet \fgt that orders even at relatively high temperatures \cite{May.2019}. The operation of a spintronic device requires ferromagnetic materials with room-temperature magnetic ordering, but most of the 2D vdW magnetic materials are either not ferromagnetic or do not possess a high ordering temperature. Here, \fgt is unique in that it has attracted specific scientific attention due to its room-temperature ferromagnetism with an ordering temperature of $T_\text{C} \approx 310 \, \text{K}$. To make use of a magnetic system for applications, spin structures are often required as information carriers. Interest in the dynamics of spin structures accordingly motivated recent studies on domain walls in vdW magnets \cite{Pei.2021, Purbawati.2020}. A particularly promising and exciting spin structure is the magnetic skyrmion \cite{Fert.2017, EverschorSitte.2018, Tokura.2021}. So for 2D magnet applications, the required next step is the observation of topological magnetic spin structures near room temperature in \fgtnospace. \fgt is additionally intriguing as it exhibits a phase transition at $\approx 100 \, \text{K}$ when it is first synthesized in a metastable phase \cite{May.2019b}. So there is particular interest in the spin structures in a \fgt crystal, before and after going through the irreversible first-order phase transition upon cooling below $T \approx 100 \, \text{K}$. These phases are termed the metastable pre-cooling and the stable post-cooling phases. Real space probing of the crystal lattice and spin structure in both of these phases is still a major open question. 

In the present work, bulk magnetic properties in the phases of 2D \fgt are characterized using a superconducting quantum interference device (SQUID), in combination with magnetic domain images using Lorentz transmission electron microscopy (L-TEM), which reveals stable topological spin structures up to room temperature. Further, the crystallographic space groups previously suggested for \fgt are investigated and we resolve open questions about their symmetry by presenting a spin chirality analysis. The spin chirality analysis involves careful considerations of the chirality of observed spin structures in \fgtnospace, in order to deduce if the chiral Dzyaloshinskii–Moriya interaction (DMI) is present in the sample. Finally, the magnetic ordering of \fgt is investigated closely and we determine whether or not the system is in a spin glassy state. This is done via a careful comparison of the effective anisotropy obtained on a macroscopic scale with the effective anisotropy obtained on a microscopic scale.

\section{Results and Discussion}

\subsection{Structural and composition analysis}
\begin{figure}[h!]
    \centering
    \includegraphics[width=\textwidth]{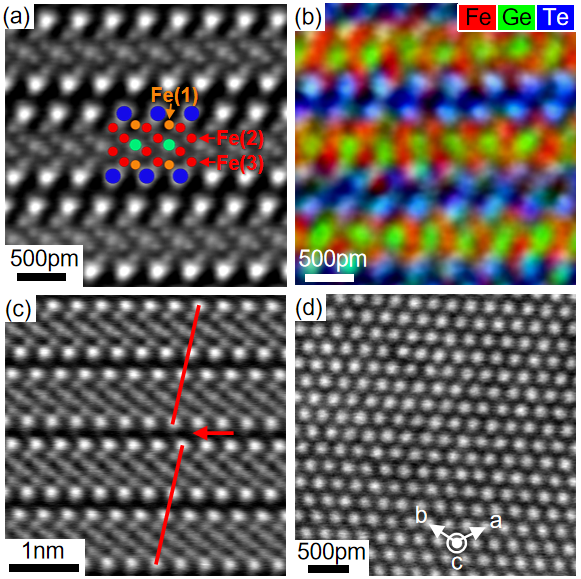}
    \caption{\textbf{High resolution analyses of the crystal structure of \fgtnospace}. (a) High resolution HAADF STEM image of the structure with the c-axis along the vertical direction in the image plane. (b) Color-coded EDX composition map showing the distribution of iron, germanium and tellurium. A schematic model of the atomic structure is overlaid on image (a) with the same color code as in panel (b). (c) HAADF STEM image showing a stacking fault as indicated by red lines and an arrow. (d) HAADF STEM image of the structure with the c-axis perpendicular to the image plane.}
    \label{crystal}
\end{figure}

Fig. \ref{crystal}(a) is a cross-section atomic resolution HAADF STEM image of the single-crystal \fgt used in the present work. Fig. \ref{crystal}(b) is a color-coded composition map obtained using EDX, which shows the periodic distribution of iron, germanium and tellurium. 2D layers consist of a sandwich of two planes of Te (brightest columns) with Ge and Fe planes located in-between. The layers are stacked along the c-axis and are approximately 1~nm thick including the vdW gap (dark area between two neighboring Te planes). A schematic model of the atomic structure is overlaid in the middle of Fig. \ref{crystal}(a) with the same color code as in Fig. \ref{crystal}(b). Following the notation used in Ref. \cite{May.2019}, the Fe(2) and Fe(3) columns can be identified but the Fe(1) columns are not visible. Since STEM images show a projection of the crystal structure over the thickness of the lamella, this iron site is not easily visible because of its split-site nature \cite{May.2019}. Fig. \ref{crystal}(c) is another HAADF STEM that shows the presence of a stacking fault as indicated by red lines and an arrow, which has been previously reported in \fgt \cite{Stahl.2018}. Here, stacking faults were observed primarily near the surface of the crystal (down to 200~nm below the surface). They are scarce deeper in the bulk of the sample, which has essentially a regular crystal structure, as shown in Fig. \ref{crystal}(a). Fig. \ref{crystal}(d) is a plan-view image where the c-axis is perpendicular to the image, which shows the hexagonal shape of the lattice.

\subsection{Magnetic imaging of skyrmionic spin structures}
We first observe the magnetization configuration at room temperature in our sample, as the type of spin structure also entails information about the underlying crystal structure of \fgtnospace. The space group of \fgt has not so far been identified unambiguously. Li et al. and May et al. report the centrosymmetric space group $R\bar3 m$ (No. 166) \cite{Li.2020, May.2019}, whereas Stahl et al. reported the non-centrosymmetric space group $R3m$ (No. 160) \cite{Stahl.2018}. Here we consider the skyrmionic spin structures in \fgt to resolve this space group issue. A prerequisite for chiral spin structures, such as skyrmions and chiral domain walls, is the DMI, which occurs in systems with broken inversion symmetry \cite{Manchon.2017}. In this regard, chiral spin structures would not be stabilized in bulk \fgt if its space group is $R\bar3m$, whereas they could be stabilized in bulk \fgt with the non-centrosymmetric space group $R3m$.  

\begin{figure}[h!]
    \centering
    \includegraphics[width=1\textwidth]{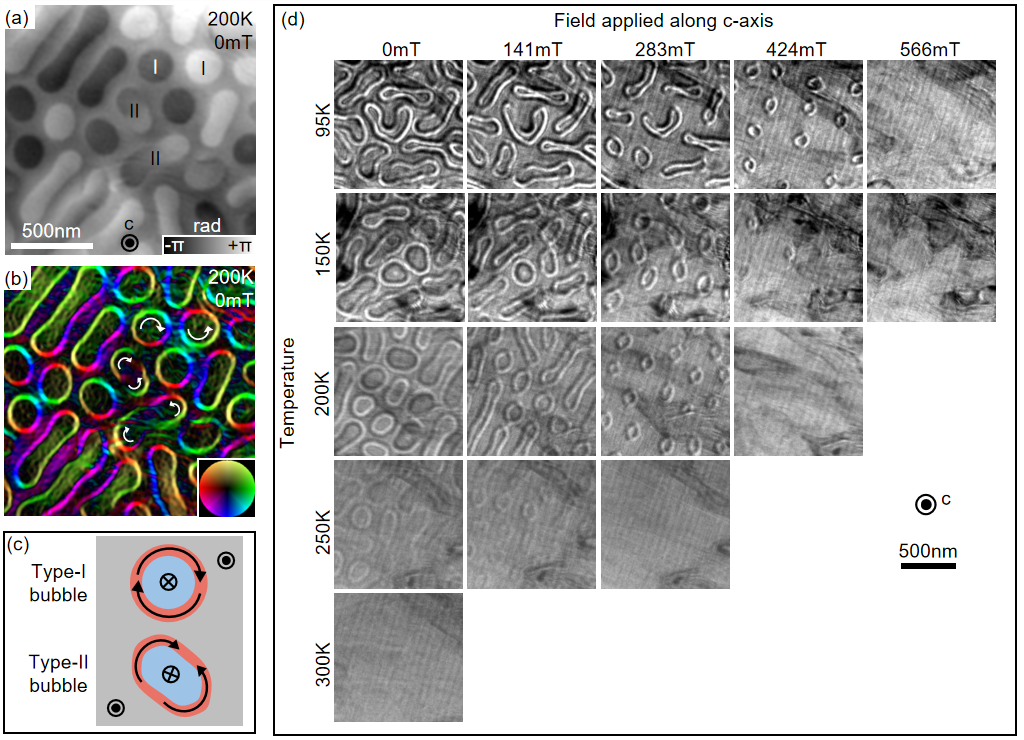}
    \caption{\textbf{Magnetic imaging of \fgtnospace, revealing skyrmionic spin structures}. (a) phase shift image obtained using off-axis electron holography at $T$ = 200 K and $B = 0 \, \text{mT}$ in \fgt with the c-axis perpendicular to the image plane after previously applying out-of-plane fields up to $B = 566 \,  \text{mT}$. (b) Corresponding color-coded magnetic induction map where the direction of the magnetic field is given by the color wheel in the bottom right corner. Type-II as well as Type-I bubbles with opposite winding numbers are present. (c) Schematic illustration of type-I and type-II bubbles. (d) Series of Fresnel images taken at various temperatures and external out-of-plane fields indicated in the figure, with a defocus of -1~mm, in a plan-view lamella (c-axis is perpendicular to the image plane). Bubbles form at higher external fields for lower temperatures.}
    \label{phase_diagram}
\end{figure}

Fig. \ref{phase_diagram}(a, b) show an in-focus phase shift image and the corresponding magnetic induction map reconstructed from an off-axis electron hologram obtained in a lamella cut perpendicular to the c-axis. The sample is magnetized along the crystallographic c-axis, which can also be deduced from SQUID measurements, as discussed in section S1 of the supplementary information. Two different types of circular spin structures can be observed. These bubbles with a black or white electron phase contrast correspond to type-I bubbles with a clockwise or counter-clockwise field rotation \cite{Loudon.2019}. These two possible chiralities were found to occur with equal probabilities. Bubbles that show a black to white phase gradient correspond to topologically trivial type-II bubbles in which the rotation of the field changes \cite{Loudon.2019}. Fig. \ref{phase_diagram}(c) shows schematically the magnetic field in type-I and type-II bubbles. The presence of both type-II and type-I bubbles with two possible chiralities indicate that they are not stabilized by DMI, but by dipolar interactions \cite{Yokota.2019}. These observations are indicative of a centrosymmetric crystal structure and thus support the results of Li et al. and May et al. regarding the centrosymmetric space group $R\bar3m$ for \fgt \cite{Li.2020, May.2019}.

A series of Fresnel defocus images taken at different temperatures and external fields, as shown in Fig. \ref{phase_diagram}(d), reveals that the formation of bubbles is favorable over labyrinth domains at larger applied fields, and that the bubbles form at fields slightly smaller than the saturating fields at any temperature. This is in line with previous results on other systems, where it was shown that skyrmion formation can be favorable at sufficiently strong external fields \cite{Lemesh.2018}. Furthermore, we observe that the sizes of the magnetic bubbles shrink as the opposing net magnetization grows with external field, which is also in line with previous reports for chiral skyrmions and non-chiral skyrmions \cite{Woo.2016, Loudon.2019}.

\subsection{Pre-cooling and post-cooling magnetic properties}

\begin{figure}[h!]
    \centering
    \includegraphics[width=\textwidth]{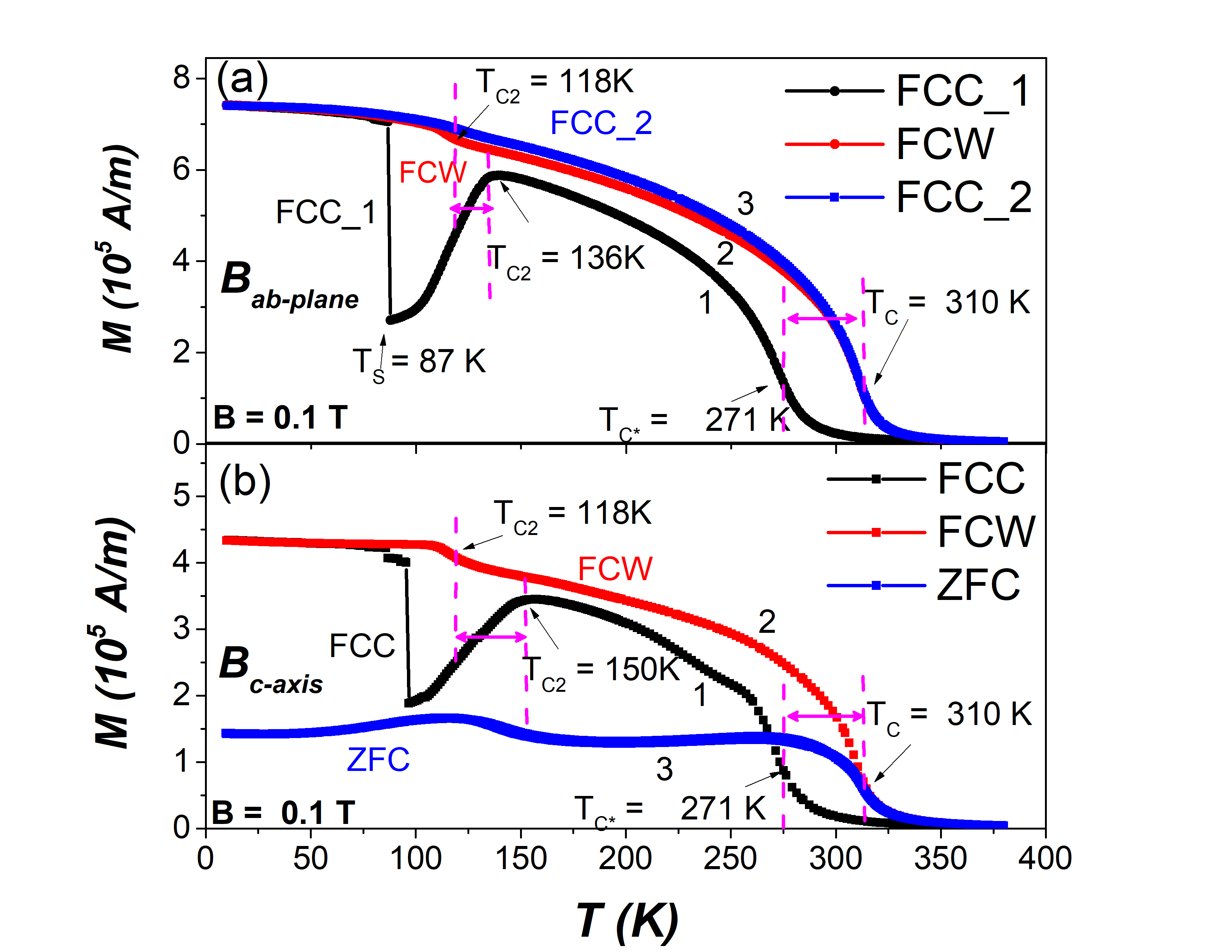}
    \caption{\textbf{Magnetization vs temperature of \fgtnospace, where the external field $B$ is applied along the c-axis and in the ab-plane with a field strength of 0.1~T.} Zero field cooled (ZFC) indicates that the sample was cooled without application of an external field and then the field of 0.1~T was applied at the lowest temperature followed by measurements performed during warming. FCC/FCW indicates measurements performed subsequently during cooling/warming with the field kept on, respectively. The numbers 1, 2, 3 defines the measurement sequence. When the external magnetic field is in the ab-pane, the sequence is FCC$_1$, followed by FCW \& finally FCC$_2$, while for $B$ along the c-axis, the sequence is FCC followed by FCW and finally ZFC.}
    \label{squidMvsT}
\end{figure}

In order to explore the bulk magnetic properties of \fgt in both the metastable pre-cooling and stable post-cooling phases, we have taken SQUID measurements of our sample while it is going through the phase transition. The temperature dependence of the magnetization for two magnetic field directions, $B$ parallel to the ab-plane ($B_\text{ab-plane}$) and $B$ parallel to the c-axis ($B_\text{c-axis}$) in an external field of 0.1~T for \fgt is shown in Fig. \ref{squidMvsT}(a) and (b), respectively. For $B_\text{ab-plane}$, the first Field Cooled Cooling (FCC$_1$) curve reveals three magnetic anomalies. The anomaly at 271~K corresponds to the magnetic ordering temperature ($T_\text{C}$). The anomaly related to decreasing magnetization at 136~K ($T_{C2}$) may correspond to a transition to a spin glassy phase \cite{Mathieu.2001}. Another possible explanation is that \fgt may be in a ferrimagnetic order under these conditions, as previously suggested for a similar temperature range by Ohta et al. and Alahmed et al., respectively \cite{Alahmed.2021, Ohta.2020}. However, as soon as our sample reaches the stable post-cooling phase, it it found to be in a ferromagnetic state. 
Finally, the anomaly of a sudden increase in magnetization at $T = 87 \, \text{K}$ corresponds to an irreversible magneto-structural phase transition ($T_S$) from a metastable pre-cooling phase to a stable post-cooling phase, consistent with previous results \cite{May.2019, Stahl.2018}. In the Field Cooled Warming (FCW) curve and the second Field Cooled Cooling (FCC$_2$) curve, no such anomaly is evident at $\approx 100 \, \text{K}$, thus confirming the 87~K anomaly of FCC$_1$ data was due to the non-reversible nature of this magneto-structural phase transition. While a weak anomaly with an increase in magnetization is evident at 118~K in FCW followed by a narrow hysteresis between FCW and FCC$_2$, it may be due to some spin reoriented phase transition. Further, the critical temperature $T_\text{C}$ has shifted from 270~K to 310~K. Such a shift in $T_\text{C}$ is consistent with previous reports \cite{May.2019b}. The difference in the net magnetization between 100~K to 300~K for FCW and FCC$_1$ curves may be partially caused by the two different spin textures that \fgt exhibits between the stable post-cooling and the metastable pre-cooling phases, which will be discussed later in more detail. 

\begin{figure}[h!]
    \centering
    \includegraphics[width=\textwidth]{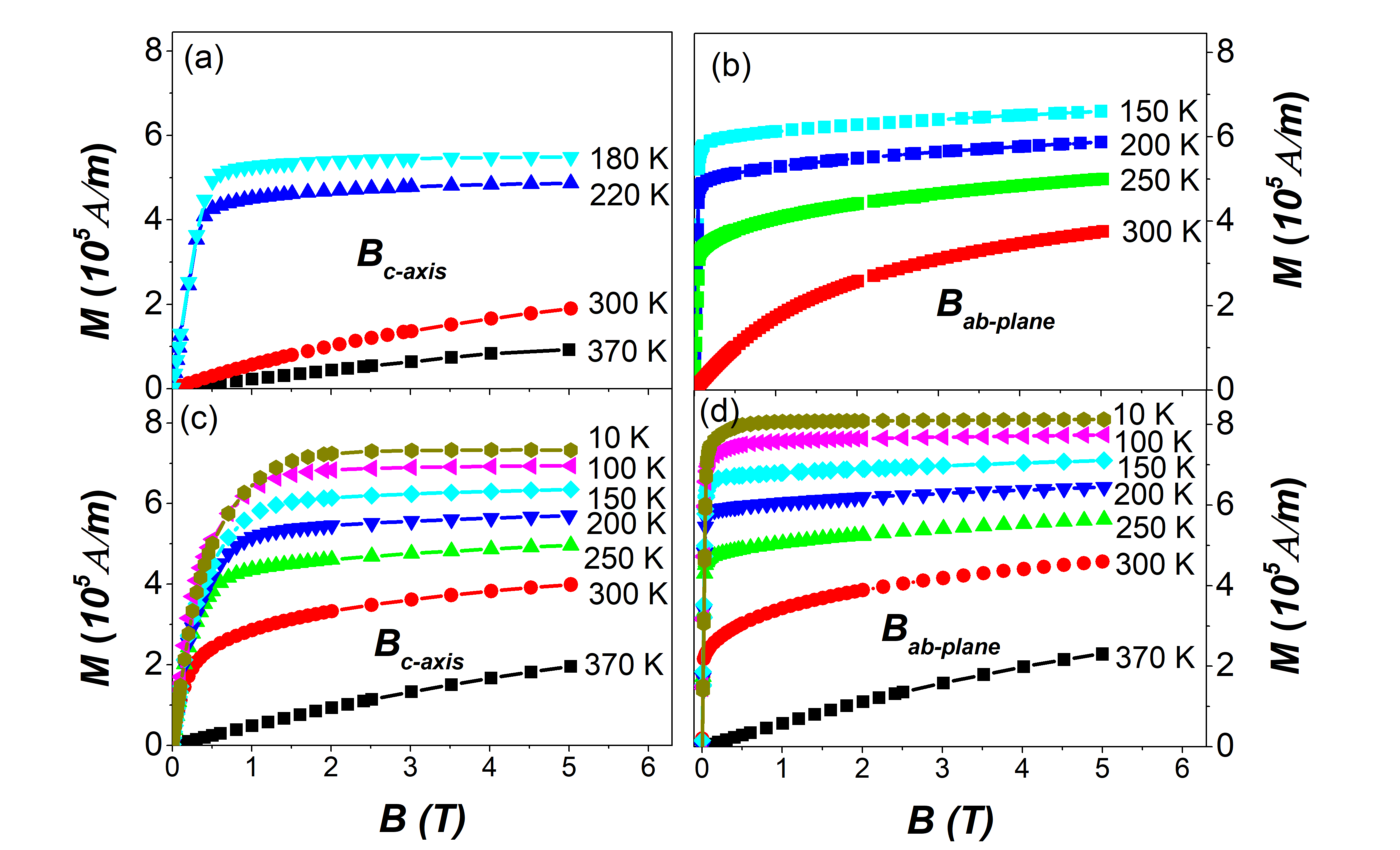}
    \caption{\textbf{The magnetic field dependent magnetization curves with the external field along the c-axis and in the ab-plane at different temperatures.} (a, b) for the pre-cooling and (c, d) for the post-cooling phase.}
    \label{squidMvsH}
\end{figure}

Due to the irreversibility of the phase transition that occurs at 87~K, newly synthesized crystal pieces were chosen for the $B_\text{c-axis}$ measurements. The $B_\text{c-axis}$ data curves show similar magnetic behavior to that of the $B_\text{ab-plane}$ configuration. The slightly different values of $T_{C2}$ and $T_S$ for the $B_\text{ab-plane}$ and $B_\text{c-axis}$ FCC curves may be due to the different magneto-crystalline anisotropy. That $T_\text{C}$ is almost the same value in both configurations further indicates that the magneto-crystalline anisotropy is a possible origin of the observed difference in $T_{C2}$ and $T_S$ for the two measurement configurations. The third and final curve is a zero-field cooled (ZFC) curve, which also shows a $T_\text{C}$ of 310~K with low net magnetization. The large difference between the ZFC and FCW curves may have resulted from the occurrence of different domain configurations due to the external magnetic field-induced reorientation of ferromagnetic components. 
In Fig. \ref{squidMvsH}, one can quantitatively see an increase in magnetization in the post-cooling phase compared to the pre-cooling phase under otherwise equivalent conditions. This is true for the wide range of measured temperatures and external fields, in both the $B_\text{c-axis}$ and $B_\text{ab-plane}$ configurations. e.g. at 150~K and 100~mT in the $B_\text{ab-plane}$ configuration:
\begin{equation}
    M_\text{pre-cooling}^\text{150K}(B = 100 \, \text{mT}) = (5.75 \pm 0.12) \cdot 10^5 \text{A m}^{-1} \; ,
\end{equation}
\begin{equation}
    M_\text{post-cooling}^\text{150K}(B = 100 \, \text{mT}) = (6.54 \pm 0.13) \cdot 10^5 \text{A m}^{-1} \; .
\end{equation}
The SQUID measurements yielded a saturation magnetization of
\begin{equation}
    M_\text{s} = (8.12 \pm 0.16) \cdot 10^5 \text{A m}^{-1} \; .
\end{equation}

To confirm the validity of this saturation magnetization value, an independent method to determine the saturation magnetization via analysis of the stripe domain patterns is used \cite{Johansen.2013}. The resulting value is $M_\text{s} = (2.56 \pm  1.40) \cdot 10^6 \, \text{A m}^{-1}$. This value has a relatively large uncertainty, and the method neglects e.g. shape anisotropy and surface defects. However, it is sufficiently close to the saturation magnetization measured with the SQUID and thus confirms its validity. The details of the method and calculation can be found in section S2 of the supplementary information.

\begin{figure}[h!]
    \centering
    \includegraphics[width=\textwidth]{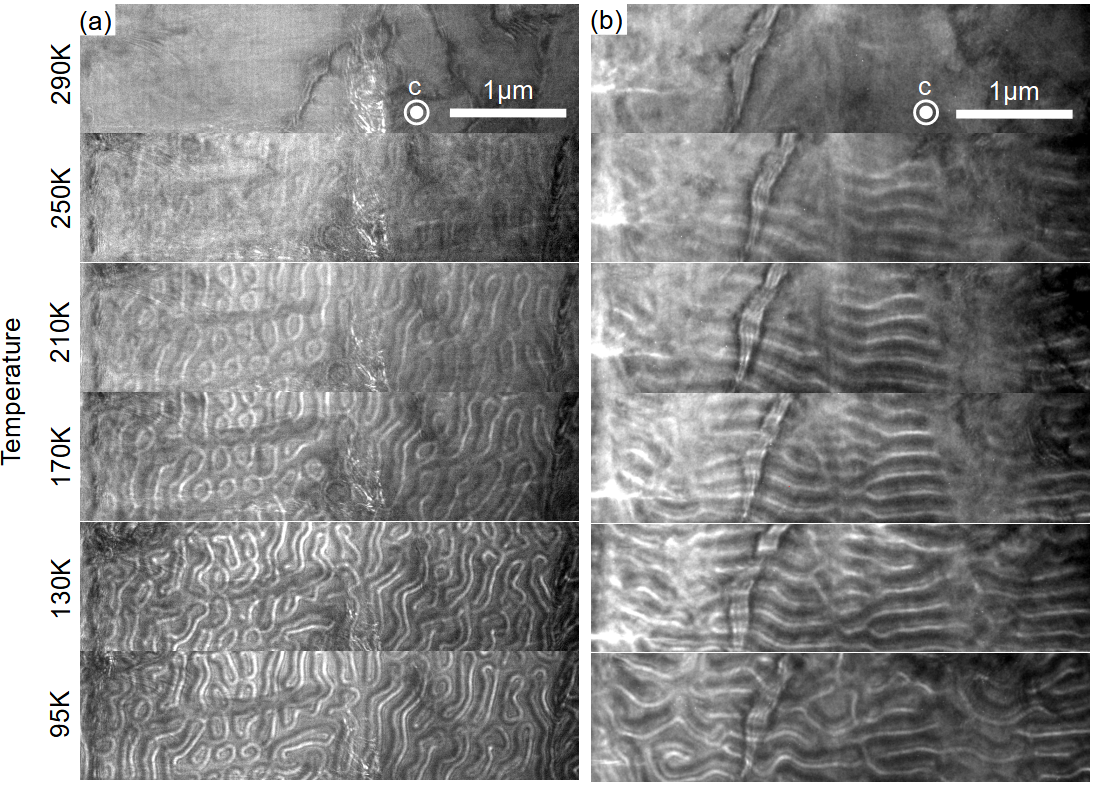}
    \caption{\textbf{Fresnel images of \fgt at various temperatures indicated in the figure and a constant external field of $B = 100 \, \text{mT}$.} A defocus of -1~mm was used in a plan-view lamella (c-axis is perpendicular to the image plane), recorded before (a) and after (b) leaving the sample in liquid nitrogen for one hour.}
    \label{tempsweep}
\end{figure}

Finally, we have taken a series of Fresnel defocus images of our sample at temperatures ranging from $95 \, \text{K}$ to $290 \, \text{K}$ and at a constant external field of $B = 100\, \text{mT}$, both in the pre-cooling and post-cooling phases. The phase transition from the metastable pre-cooling to the stable post-cooling phase in the TEM lamellas was achieved by cooling the sample down to $\approx 77 \, \text{K}$, leaving it in liquid nitrogen for one hour. The images for these temperature sweeps are shown in Fig. \ref{tempsweep}. A striking difference between the two sets of images is that skyrmions are formed in the pre-cooling phase, but stripe domains form in the post-cooling phase, under otherwise equivalent circumstances. Since magnetic skyrmions predominantly form when an external field close to the saturation field is applied, this suggests that the saturation field of our sample has increased during the phase transition to the post-cooling phase. This could be explained by an increase in the magnetization in the post-cooling state, as this would mean that the dipolar energy is still large relative to the external field contribution. Magnetometry measurements of a bulk sample using a SQUID support this idea, and yield larger magnetizations at moderate fields in the post-cooling phase compared to the pre-cooling phase, as discussed earlier in this subsection.

\subsection{Magnetic ordering}
It was shown in the previous subsection that \fgt shows a decrease in magnetization upon cooling from $T_{C2}$ down to $T_S$ while the sample is in the metastable pre-cooling phase (see Fig. \ref{squidMvsT}), which indicates that \fgt might exhibit ferrimagnetic order or is in a spin glassy state. To distinguish these effects, we study next whether or not \fgt is in a glassy spin state. To this end, the micromagnetic anisotropy constant $K$ is experimentally determined in two independent ways and compared. One method is to consider the hard-axis saturation field $B_\text{sat}^\text{IP}$, which is simply related to the exchange constant via the Stoner-Wohlfarth model \cite{Stoner.1948}. By finding the total energy minimum with a given field applied along the hard-axis, and including Zeeman and anisotropy interactions, one finds that the expression for in-plane saturation is:
\begin{equation}
    \label{K_IP}
    K_\text{IP} = B_\text{sat}^\text{IP} \frac{M_\text{s}}{2} \; .
\end{equation}
A complementary approach to determine $K$ is via the domain wall width $\delta$ \cite{Heide2008a}:
\begin{equation}
    \label{KDW_eq}
    \delta = \pi \Delta = \pi \sqrt{\frac{A}{K_\delta}} \; ,
\end{equation}
where $A$ is the micromagnetic exchange constant and $\Delta$ is a fit parameter used in the determination of the domain wall width $\delta$ from the measured domain wall profile. Note that when measuring $K_\text{IP}$, the quantities involved are determined on a macroscopic scale, whereas the determination of $K_\delta$ will require knowledge of the exchange interaction, which is short-ranged and dominates on the nanoscale. Thus, local disorder present in spin glasses \cite{Sherrington.1975} would manifest itself by affecting $K_\delta$ and $K_\text{IP}$ differently. 


To obtain $K_\delta$, the micromagnetic exchange coupling constants were calculated using the JuKKR density functional theory (DFT) code \cite{jukkr} (see method section \ref{subsec:DFT}). This code employs the method of infinitesimal rotations \cite{Liechtenstein1987,Schweflinghaus2016} to map the interactions among spins $\vec{S}_i=\vec{M}_i/\mu_i$ ($\mu_i=|\vec{M}_i|$) at lattice sites $i$ onto the extended classical Heisenberg Hamiltonian
\begin{equation}
    H = - \sum_{i \neq j} J_{ij} \big(\vec S_i \cdot \vec S_j \big) - \sum_{i \neq j} \vec D_{ij} \cdot \big(\vec S_i \times \vec S_j \big) - \sum_{i} K (\vec S_i \cdot \hat e_z)^2 \, ,
\end{equation}
where the first term describes the exchange interaction, the second term represents DMI, and the final term describes the uniaxial out-of-plane anisotropy. The micromagnetic spin stiffness is 
\begin{equation}
    A = \frac{1}{2V} \sum_{i \neq j} J_{ij} \vec r_{ij}^2 \, ,
    \label{A_DFT_equation}
\end{equation}
where $V$ is the volume per Fe atom, which can be calculated from the exchange coupling constants \cite{Schweflinghaus2016}. These parameters define the micromagnetic energy functional 
\begin{equation}
E[\vec m] = \int d \vec r \left[  A \dot {\vec m}^2 + Km^2_z \right], 
\end{equation}
where the contribution of the spiralization that is related to the DMI vectors is neglected because it is found to vanish.
The micromagnetic exchange constant obtained from this method is
\begin{equation}
    A_\text{DFT} = 0.1508 \; \text{eV nm}^{-1} \; . 
\end{equation}
An alternative, independent method to approximate the micromagnetic exchange constant $A$ by considering the magnetic ordering temperature and crystal structure, is presented and executed in section S3 of the supplementary information. The result is $A_\text{H} = (0.060 \pm 0.007) \, \text{eV nm}^{-1}$ and it is fairly close to the DFT value considering that many assumptions are made in this alternative method. Further calculations and comparisons to experiment, which support the DFT models and calculations are presented in the supplementary sections S4 and S5.

\begin{figure}[h!]
    \centering
    \includegraphics[width=\textwidth]{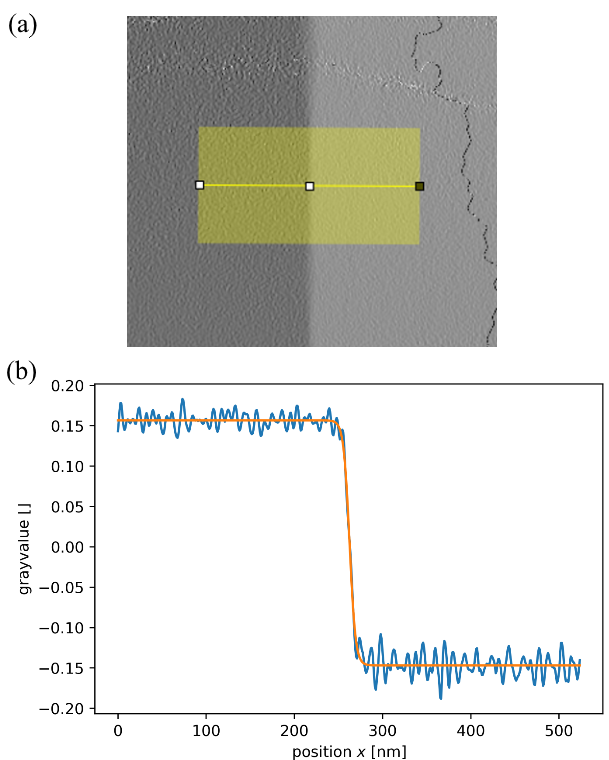}
    \caption{ \textbf{Analysis of the magnetic contrast of a magnetic domain wall, with the aim to determine the domain wall width.} (a) Magnetic phase gradient image obtained using off-axis electron holography. A profile was extracted in the direction perpendicular to a 180° domain wall along the wide line. (b) Phase gradient profile (blue) and corresponding fit (orange). This example shows one magnetic profile across a domain wall at 95~K. }
    \label{KDW}
\end{figure}

To determine the effective anisotropy $K_\delta$ using the spin stiffness $A$, we next need to determine the domain wall width $\delta$. The domain wall parameter $\Delta$ can in principle be determined from phase shift images obtained using off-axis electron holography by fitting the magnetic phase gradient profile perpendicular to a domain wall and using the well-known theoretical profile $M_\text{z}^\text{theo}(x) = M_0 \cdot \tanh(x/\Delta)$. In reality the profiles will not be perfectly centered, so we use a fit function which allows for shifts:
\begin{equation}
    M_\text{z}^\text{fit}(x) = M_0 \cdot \tanh((x - x_\text{shift})/\Delta) + M_\text{shift} \; ,
\end{equation}
where $M_0$ is the amplitude of the magnetization, $x$ is the position perpendicular to the domain wall, and $x_\text{shift}$ and $M_\text{shift}$ are the free fit parameters accounting for imperfect centering. To obtain reliable results, each line drawn perpendicular to the domain walls has a width of 277.5~nm (300 pixels) to average across. An example of such a line is shown in Fig. \ref{KDW}(a), and the corresponding magnetic profile and fit is shown in panel (b). Overall, the domain wall width has been determined via four independent lines at $95 \, \text{K}$ and five different lines at $200 \, \text{K}$. The resulting domain wall widths are $\delta^{95\text{K}} = (24.2 \pm 5.2) \, \text{nm}$ and $\delta^{200\text{K}} = (36.3 \pm 3.5) \, \text{nm}$. The uncertainties are estimated via the standard deviation of the domain wall widths obtained from all lines used at each temperature. Accordingly, the fit parameters are $\Delta = \delta/\pi$. Thus, all required parameters to find $K_\delta$ via equation \ref{KDW_eq} are known, and we find using $A_\text{DFT}$ and the parameters $\Delta$:
\begin{equation}
    \label{KDW_num95}
    K_\delta^\text{95K} = (2.55 \pm 0.55) \cdot 10^{-3} \, \text{eV nm}^{-3} \; ,
\end{equation}
\begin{equation}
    \label{KDW_num200}
    K_\delta^\text{200K} = (1.13 \pm 0.11) \cdot 10^{-3} \, \text{eV nm}^{-3} \; .
\end{equation}

For the alternative determination of $K$ via the saturation field along the hard-axis, images of the cross-section lamella sample are recorded at various field strengths at $95 \, \text{K}$ and $200 \, \text{K}$ (examples of images at $95 \, \text{K}$ are shown in Fig. \ref{fresnel_cinplane}(c-e)). 
The point at which the magnetic contrast completely vanishes as a function of magnetic field is approximated by splining the data for which there is magnetic contrast, and interpolating to the point where the minority domain has a width of zero. It should be noted that, since the field may not be applied perfectly perpendicular to the c-axis, this field is the monodomainization field, and serves as a lower bound for the hard-axis saturation field. However, it can be used as an estimation for the hard-axis saturation field. In this case, the saturation fields are estimated to be $B_\text{sat}^\text{IP, 95K} = (285 \pm 85) \, \text{mT}$ and $B_\text{sat}^\text{IP, 200K} = (143 \pm 43) \, \text{mT}$. 
Accordingly, in conjunction with the saturation magnetization $M_\text{s}$ from the SQUID measurements and inserting into equation \ref{K_IP}, the micromagnetic anisotropy constants from this method are evaluated to be
\begin{equation}
    K_\text{IP}^{\text{95K}} = (0.72 \pm 0.22) \cdot 10^{-3} \, \text{eV nm}^{-3} \; ,
\end{equation}
\begin{equation}
    K_\text{IP}^{\text{200K}} = (0.36 \pm 0.11) \cdot 10^{-3} \, \text{eV nm}^{-3} \; .
\end{equation}


Finally, the anisotropy constants obtained via both the domain wall and hard-axis saturation method can be compared. At 95 K the ratio between the effective anisotropy constants is $K_\delta^\text{95K}/K_\text{IP}^\text{95K} = 3.53 \pm 1.30$, and at 200 K it is $K_\delta^\text{200K}/K_\text{IP}^\text{200K} = 3.11 \pm 0.98$. Thus, the two methods to determine $K$ used in this work, although they are completely independent and cover vastly different length scales, yield results which are on the same order of magnitude. The fact that the ratio $K_\delta/K_\text{IP}$ may systematically be larger than $1$ could be attributed to the fundamental approximations used. Namely, since the hard-axis saturation field will be slightly larger than the monodomainization field, which we used to estimate the saturation field, $K_\text{IP}$ should in reality be systematically slightly larger than the values we obtained. If the obtained anisotropy constants would differ by a large margin one could suggest that the material might be in a spin glassy state. But since they do not differ greatly, this analysis implies that the material is not in a spin glassy state. As such, this analysis in conjunction with the SQUID results indicates that \fgt might not be in a spin glassy state, but in a ferrimagnetic state while it is in the pre-cooling phase, as previously claimed  \cite{Alahmed.2021, Ohta.2020}. 

\section{Conclusions}
In conclusion, we have presented experimental observations of spin structures and magnetic properties in the promising 2D vdW magnet \fgtnospace, to understand the magnetic ordering of this material. Notably, non-chiral magnetic skyrmions form in the material near room temperature when magnetic fields close to the saturation field are applied. The achirality of these magnetic bubbles suggests that there is no considerable DMI present in \fgtnospace, which in turn supports the conclusion that \fgt has the centrosymmetric $R \bar3 m$ crystal structure, rather than the otherwise proposed non-centrosymmetric $R 3 m$ crystal structure. Since \fgt exhibits a lower magnetization in the pre-cooling phase than in the post-cooling phase, we were able to observe bubbles at lower fields in the pre-cooling phase than in the post-cooling phase. Furthermore, by considering the anisotropy constants obtained via two independent methods, namely via the hard-axis saturation field and via the domain wall profile and predicted exchange constants, we conclude that \fgt is not in a spin glassy state. However, a notable drop of the magnetization with decreasing temperature observed in SQUID measurements suggests that \fgt is a ferrimagnet instead of a simple ferromagnet whilst in the pre-cooling phase. Future neutron diffraction experiments are required to fully ascertain the spin structure of \fgtnospace. We have thus completed the necessary steps toward room temperature spintronics devices based on Skyrmion type spin structures, which are of significant interest for novel information storage and non-conventional computing concepts using 2D magnets.

\section{Methods}

\subsection{A. Crystal synthesis}
Single crystals of \fgt with an average size of $2 \times 2 \times 0.5 \, \text{mm}^3$ were synthesized in quartz glass ampoules from the elements Fe, Ge, and Te in a 6 : 1 : 2 ratio in the presence of iodine as mineralizer. Similar to reference \cite{May.2019}, Ge was supplied as a powder on the bottom of the quartz glass tube. Fe and Te were pressed to separate pellets of $1 \, \text{mm}$ diameter and positioned on top of the Ge powder without direct contact between the pellets. The reaction thus predominantly occurs through the gas phase and increases the single crystal size to several millimeters. The vacuum-sealed ampules were heated up to 750 °C with 120 K/h and kept at this temperature for about 2 weeks before quenching them in room temperature water. The composition of the single crystals was confirmed by energy dispersive X-ray spectroscopy (Tescan SEM Vega TS 5130 MM equipped with a silicon drift detector, Oxford) with a ratio of Fe : Ge : Te of 4.72(5) : 1 : 1.94(4). 

\subsection{B. Lamella preparation}
Electron transparent cross-section and plan-view lamellas were prepared using a 30~kV Ga$^{+}$ focused ion beam and scanning electron microscope (FIB-SEM) FEI Helios platform. The energy of the ion beam was decreased to 5~kV in the last steps of the thinning to minimize surface damage.

\subsection{C. Transmission electron microscopy}
Scanning transmission electron microscopy (STEM) and energy dispersive X-ray (EDX) spectroscopy were carried out using an FEI Titan TEM equipped with a Schottky ﬁeld emission gun operated at 200~kV, a CEOS probe aberration corrector, a high angle annular dark-ﬁeld detector (HAADF) and a Super-X EDX detection system \cite{Kovacs2016}. Composition maps were obtained using the Thermo Scientific Velox software. 

Magnetic imaging was carried out using Fresnel defocus and off-axis electron holography in an FEI Titan TEM equipped with a Schottky field emission gun, a CEOS image aberration corrector, a post-specimen electron biprism and a 4k$\times$4k Gatan K2-IS direct detection camera \cite{Boothroyd2016}. A liquid-nitrogen-cooled specimen holder (Gatan model 636) was used to vary the sample temperature. The microscope was operated at 300~kV in magnetic field-free conditions (Lorentz mode) by using the first transfer lens of the aberration corrector as the primary imaging lens. The conventional objective lens was used to apply the chosen magnetic fields to the sample, which were pre-calibrated using a Hall probe. The field is applied along the electron beam direction \textit{i.e.} perpendicular to the sample plane. For off-axis electron holography, the electron biprism was used to overlap a reference wave travelling in vacuum with an object wave passing through the sample. An elliptical illumination was used to optimize the coherence of the beam in the direction perpendicular to the biprism. The hologram width was 2.7~µm and the fringe spacing was 2.5~nm. Magnetic induction maps were reconstructed using Fourier transforms with the Holoworks plugin in the Digital Micrograph software (Gatan) \cite{VOLKL.1995}. The contribution of the mean inner potential to the phase was removed by subtracting the phase of an hologram acquired above the Curie temperature.

\subsection{D. DFT atomistic exchange tensor calculations} 
\label{subsec:DFT}
We performed density functional theory (DFT) calculations for bulk \fgt using the experimental lattice constants, $a = 4.04(2)$\AA, $c = 29.19(3)$\AA, from Ref. \cite{May.2019}. In order to account for the two Fe atoms at 50\% occupancy we employed the use of the coherent potential approximation (CPA) \cite{Ebert2011a}. We perform our calculations with the JuKKR code \cite{jukkr} which implements the full potential Korringa Kohn Rostoker (KKR) Green's Function method \cite{Ebert2011a} with an exact description of the shape of the atomic cell \cite{Stefanou1990, Stefanou1991}. We use an angular momentum cutoff $l_{max} = 3$ and a generalised gradient approximation (GGA) to obtain the exchange-correlation potential \cite{Perdew1996}. Our calculations of the exchange coupling constants for the determination of the atomistic spin stiffness are based on the model of infinitesimal rotations \cite{Liechtenstein1987,Schweflinghaus2016}. The domain wall widths calculated in the supplementary S5 are determined using the \textit{Spirit} code \cite{Muller2019}.

\section{Data Availability}
All data used in the present work are available upon reasonable request.

\section{Code Availability}
All code used in the present work is available upon reasonable request.

\printbibliography

\section{Acknowledgements}
The work at JGU Mainz was funded by the Deutsche Forschungsgemeinschaft (DFG, German Research Foundation) - TRR 173 -  268565370 (projects A01 and B02), project 403502522 (SPP 2137 skyrmionics) the EU (FET-Open grant agreement no. 863155 (s-Nebula)), and the Research Council of Norway (QuSpin Center 262633). The work at JGU Mainz and FZJ was funded by ERC Synergy grant agreement no. 856538 (3D MAGiC), T.S., P.R., S.B. and Y.M. gratefully acknowledge the Jülich Supercomputing Centre for providing computational resources and Deutsche Forschungsgemeinschaft (DFG, German Research Foundation) - TRR 173 - 268565370 (project A11), TRR 288 - 422213477 (project B06). T.G.S. Would like to thank Dr. Sarah Jenkins and Dr. Fabian Lux for fruitful discussions.

\section{Author Information}
\subsection{Affiliations}
\textbf{Johannes Gutenberg Universität Mainz, Institut für Physik, Staudingerweg 7, 55128 Mainz, Germany}
\newline
Maurice Schmitt, Tom G. Saunderson, Aga Shahee \& Mathias Kläui
\newline \newline
\textbf{Ernst Ruska Centre for Microscopy and Spectroscopy with Electrons and Peter Grünberg Institute, Forschungszentrum Jülich, 52425 Jülich, Germany}
\newline
Thibaud Denneulin, András Kovács, Amir Tavabi \& Rafal Dunin-Borkowski 
\newline \newline
\textbf{Forschungszentrum Jülich GmbH, Wilhelm-Johnen-Straße, 52428 Jülich, Germany}
\newline
Philipp Rüßmann, Yuriy Mokrousov \& Stefan Blügel
\newline \newline
\textbf{Max Planck Institute for Solid State Research, Heisenbergstraße 1, 70569 Stuttgart, Germany}
\newline
Tanja Scholz, Bettina Lotsch
\newline \newline
\textbf{University of Bristol, School of Physics, HH Wills Physics Laboratory, Tyndall Avenue, Bristol BS8 1TL}
\newline
Martin Gradhand
\newline\newline
\textbf{National and Kapodistrian University of Athens, Department of Physics, University Campus, GR-157 84 Zografou, Athens}
\newline
Phivos Mavropoulos

\subsection{Contributions}
Tanja Scholz synthesized all \fgt samples used in the present work. Thibaud Denneulin prepared the TEM lamellas and conducted the LTEM experiments with the help of Amir Tavabi. András Kovács carried out the STEM and EDX measurements. Aga Shahee conducted the SQUID measurements. Tom G. Saunderson and Philipp Rüßmann contributed equally to the DFT results. Martin Gradhand and Phivos Mavropolous helped with the determination of the spin stiffness via $T_C$. Phivos Mavropoulos provided the theoretical $T_C$ of \fgt based on DFT results. Maurice Schmitt analyzed the experimental and theoretical results to yield insights into the spin ordering of \fgtnospace, and wrote the paper with Mathias Kläui. Bettina Lotsch supervised the synthesization of the \fgt crystals. Yuriy Mokrousov and Stefan Blügel discussed and supervised the DFT work. Rafal Dunin-Borkowski supervised the microscopy work. Mathias Kläui supervised the experimental SQUID measurements and devised the study. All authors commented on the results and contributed to the manuscript.

\section{Ethics declarations}
\subsection{Conflict of Interest}
The Authors declare that there is no conflict of interest.

\newpage\newpage

\section{Supplementary information}
\subsection{S1 Magnetic anisotropy in \fgt}
Isothermal magnetization vs. magnetic field (M vs. H) curves measured using a SQUID, under the applied field either parallel to the c-axis ($B_\text{c-axis}$) or the ab plane ($B_\text{ab-plane}$) are shown in Fig. \ref{squidMvsH} of the main text. Fig. \ref{squidMvsH}(a) and (b) are for the pre-cooling phase and Fig. \ref{squidMvsH}(c) and (d) for the post-cooling phase respectively. These magnetization curves are presented without taking into consideration the demagnetization factor, which plays a great role in defining the internal field in 2D flake-like and 1D needle-like crystals. These M vs. H curves indicate a soft ferromagnetic nature of both the pre-cooling and post-cooling phases with a weak magneto-crystalline anisotropy. The raw data indicates a smaller external field is needed to saturate magnetization in the ab-plane than along the c-axis, which may at first indicate that \fgt is an easy plane ferromagnet. However, consideration of the large contribution to internal field from the demagnetization factor for the c-axis configuration indicates an opposite behavior, where spins prefer to align along the c-axis, as has been reported by May et al. \cite{May.2019}.

\begin{figure}[h!]
    \centering
    \includegraphics[width=\textwidth]{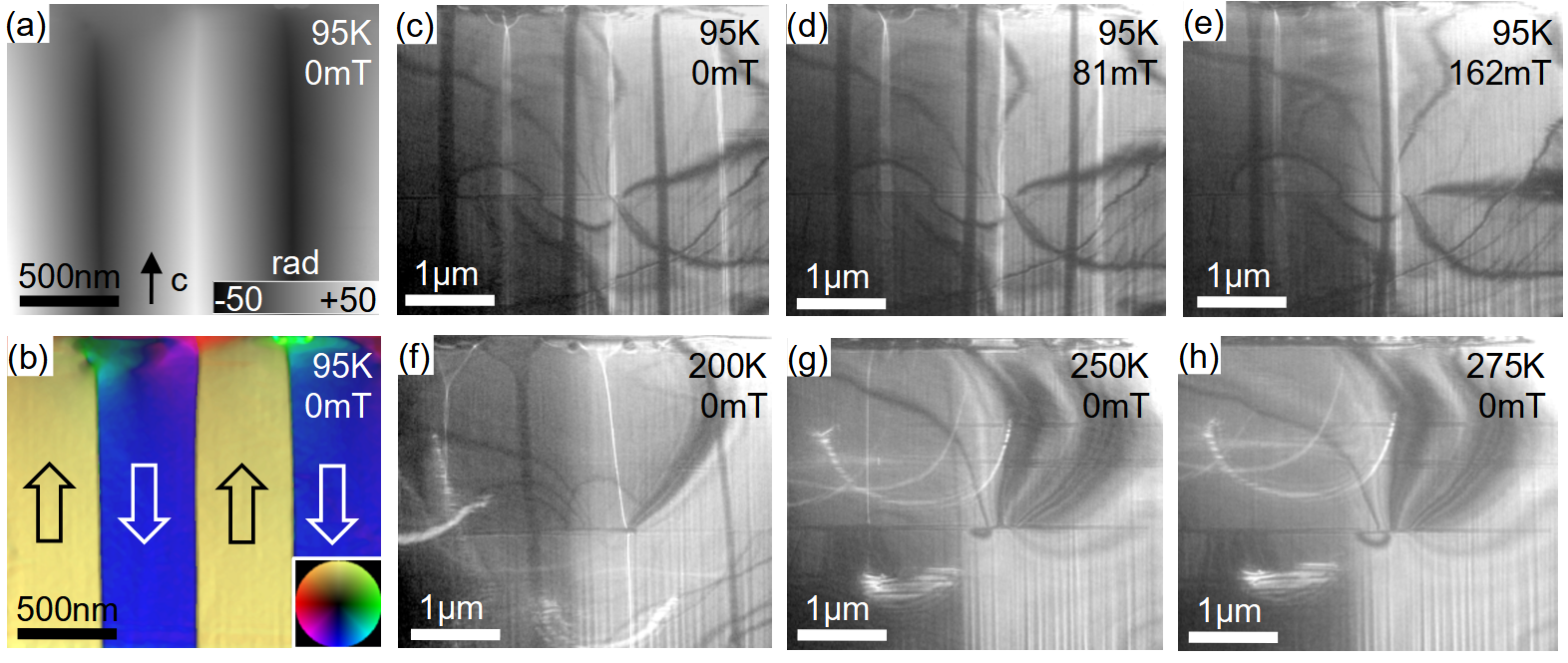}
    \caption{\textbf{L-TEM images of a \fgt lamella with the crystallographic c-axis in the image plane.} (a) phase shift image obtained using off-axis electron holography at 95~K and 0~mT in a cross-section lamella (c-axis is vertical in the image plane). (b) Corresponding color-coded magnetic induction map where the direction of the magnetic field is given by the color wheel in the bottom right corner. (c-e) Fresnel images obtained at 95~K and different external fields indicated on the images and with a defocus of -1~mm. (f-h) Fresnel images obtained at 0~mT and different temperatures indicated on the images.}
    \label{fresnel_cinplane}
\end{figure}

Further, we investigated magnetic domains in lamellas cut from \fgt crystals parallel to  the c-axis. Fig. \ref{fresnel_cinplane}(a,b) shows a phase shift image and the corresponding color-coded magnetic induction map obtained at 95~K and 0~mT. The presence of 180° stripe domains with domain walls (DWs) orientated along the [001] direction confirm the easy-axis magnetic anisotropy of \fgt with the c-axis (001) as the easy-axis. Fig. \ref{fresnel_cinplane}(c-h) shows Fresnel defocus images obtained in the presence of different external fields applied nearly perpendicular to the c-axis (0 to 162~mT) and at different temperatures (95 to 275~K). Even though the domain walls move in the presence of external fields and the width of the domains changes as a function of temperature, it can be observed that the DWs remain parallel to the c-axis even close to $T_\text{C}$ at 250~K for instance, which confirms the easy-axis magnetic anisotropy of \fgt over a wide range of temperatures, which as shown in the main text, we have observed directly using L-TEM.

\subsection{S2 Determination of the saturation magnetization from the stripe domain pattern}
\begin{figure}[h!]
    \centering
    \includegraphics[width=\textwidth]{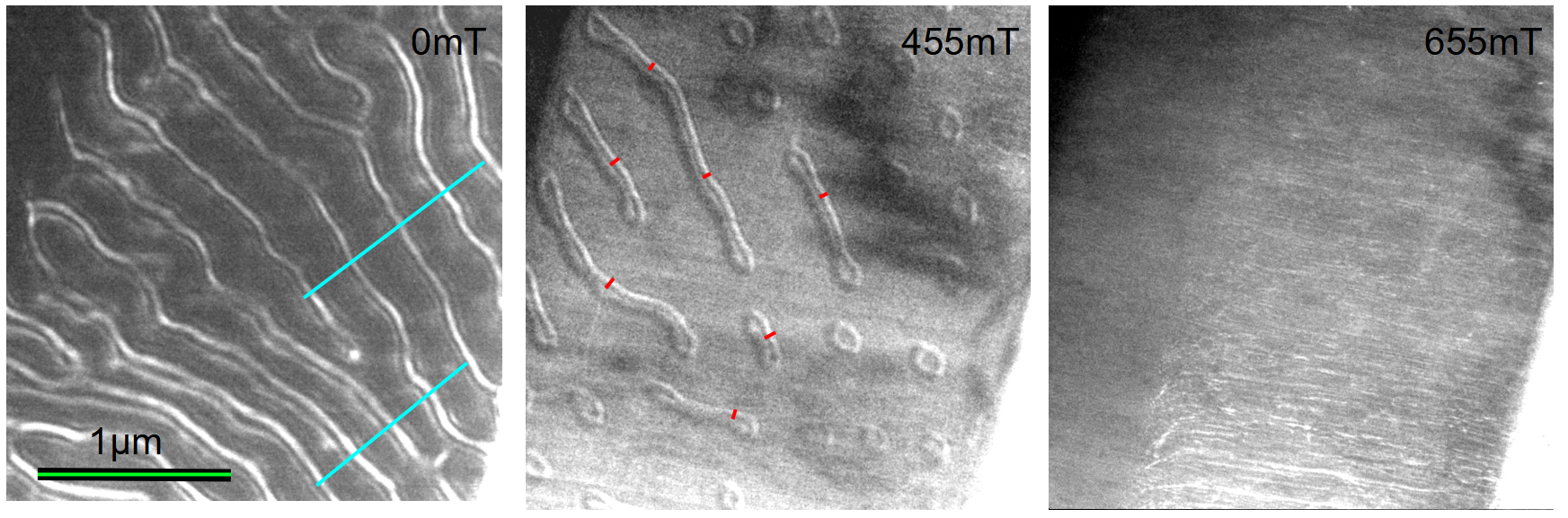}
    \caption{\textbf{L-TEM images of \fgt at $95 \, \text{K}$.} The red lines are used to estimate the minority stripe domain width, whereas the blue lines are used to determine the overall domain periodicity. The green line is used to convert lengths in pixels to real lengths. The c-axis is perpendicular to the image plane.}
    \label{johansenTEM}
\end{figure}

An independent way to determine the saturation magnetization is to closely examine the stripe domain patterns seen in the L-TEM images. This way, the validity of the SQUID results can be checked. According to theoretical considerations \cite{Johansen.2013}, the saturation magnetization of a film of magnetic material is given by
\begin{equation}
    \label{johansenMs}
    M_\text{s} = 2 \cdot \frac{a^2}{t} \cdot \frac{(B_\text{c} - B)/\mu_0}{\pi r_{\downarrow \text{c}}} \; ,
\end{equation}
where $\mu_0$ is the magnetic vacuum permeability, $M_\text{s}$ is the saturation magnetization, $a$ is the overall domain periodicity, $t$ is the film thickness, $r_{\downarrow \text{c}}$ is the width of the minority spin domains, and $B$ and $B_\text{c}$ are an arbitrarily chosen field and the saturating field, respectively. This theory assumes the considered sample is a magnetic film with uniaxial anisotropy and a large domain period, and considers exchange, Zeeman, anisotropy, and stray field energy contributions. The film thickness in our sample is not uniform, as scanning electron microscopy (SEM) measurements on various spots of the sample reveal. 
However, the measured local thicknesses all range from $90 \, \text{nm}$ to $210 \, \text{nm}$, so the film thickness can be estimated to be $t = (150 \pm 60) \, \text{nm}$. All other parameters on the right hand side of equation \ref{johansenMs} can be estimated from the recorded TEM images. The domain periodicity $a$ at the chosen field $B = 0 \, \text{mT}$ can be estimated by adding the length of both blue lines in Fig. \ref{johansenTEM} and dividing by eight, since both blue lines together span a total of eight domain periods. The average length of the red lines is used to estimate $r_{\downarrow \text{c}}$. 
Since no magnetic contrast is visible at $655 \, \text{mT}$ anymore, the saturating field is estimated to be $B_\text{c} = (550 \pm 100) \, \text{mT}$ in this case. This method leads to a relatively large uncertainty in the minority domain width, since it is not only a relatively small distance to measure, but also because the measured widths cannot be measured exactly at the critical field. Therefore, each individual red line can be estimated to have an error of $50 \, \%$, which propagates to yield the average length of all lines of $r_{\downarrow \text{c}} = (53.6 \pm 10.2) \, \text{nm}$. Via the same method, the domain periodicity is estimated to be $a = (272.1 \pm 96.5) \, \text{nm}$. 

Accordingly, using error propagation, the obtained saturation magnetization is $M_\text{s} = (2.56 \pm  1.52) \cdot 10^6 \, \text{A m}^{-1}$. Although this way the saturation magnetization has a larger relative uncertainty, its order of magnitude agrees with the values obtained from the SQUID measurements, thus confirming the validity of the SQUID measurements.

\subsection{S3 Determination of the micromagnetic exchange constant from the Curie temperature}
To confirm the results obtained from DFT, the micromagnetic exchange constant (or spin stiffness) will be approximated in the following. To this end, one can invoke a simple Heisenberg model, which predicts in a mean-field approach \cite{Evans.2014}: 
\begin{equation}
\label{heisenbergmean}
    J_\text{H} =  \frac{3 k T_\text{C}}{2 \epsilon z} 
\end{equation}
where $k$ is the Boltzmann constant, $T_\text{C}$ is the Curie temperature, $z$ is the number of nearest neighbors in a unit cell and $J_\text{H}$ is the Heisenberg exchange constant between magnetic sites. The parameter $\epsilon$ is a correction factor accounting for spin waves, which is slightly smaller than one and is well-known for common crystal structures \cite{Garanin.1996}. Here, we will simply estimate $\epsilon = 0.8 \pm 0.1$. One can then find the micromagnetic exchange constant $A_\text{H}^\text{IP}$ 
according to equation \ref{A_DFT_equation} by assuming $J_{ij} = J_H$ for nearest neighbors only and averaging the sum over nearest neighbors as well as the nearest neighbor distance squared $\overline{r_{ij}^2}$:
\begin{equation}
    A_\text{H}^\text{IP} = J_H \cdot \frac{z \cdot \overline{r_{ij}^2}}{2 V} \; ,
    \label{micromagneticA}
\end{equation}
with the volume per iron atom $V$.
Combining \ref{heisenbergmean} and \ref{micromagneticA} yields 
\begin{equation}
\label{A_DW}
    A_\text{H}^\text{IP} = \frac{3 k T_\text{C}}{4 \epsilon} \cdot \frac{\overline{r_{ij}^2}}{V} \; ,
\end{equation}
The Curie temperature $T_\text{C} = (310 \pm 10) \, \text{K}$ can be estimated from the SQUID data. Note that in \fgt the Curie temperature changes depending on the thermal phase it is in. The larger value of $T_\text{C} = (310 \pm 10) \, \text{K}$ is valid for the post-cooling phase. 
The volume per magnetic site is $V = 0.0276 \, \text{nm}^3$ according to the DFT results. For the mean squared neighbor distance, all neighbors in table \ref{neighbortable} are considered and averaged over. Here, the split site nature of the Fe$_1$ site has to be considered, by weighting them only $50 \%$. The result is $\overline{r_{ij}^2} \approx 0.06594 \, \text{nm}^2$. Inserting these values into equation \ref{A_DW} yields
\begin{equation}
    A_\text{H}^\text{IP} = (0.060 \pm 0.007) \, \text{eV nm}^{-1} \; .
\end{equation}
The value of the micromagnetic exchange constant obtained this way $A_\text{H}$ is close to the value obtained via DFT $A_\text{DFT}$. Considering that many assumptions are made when calculating the micromagnetic exchange constant via the critical temperature due to the complex crystal structure of \fgtnospace, the obtained value $A_\text{H}$ turns out to be compatible with the more robust DFT result $A_\text{DFT}$ and thus supports it.

\begin{table}[h!]
    \centering
    \begin{tabular}{|c|c|c|c|}
    \hline
        type & Neighbor & Number of neighbors & Bond length \\
        \hline
        Fe$_1$ & Fe$_3$ & 3 & 2.35464 \AA \\
        \hline
        Fe$_1$ & Fe$_2$ & 3 & 2.73786 \AA \\
        \hline
        Fe$_2$ & Fe$_3$ & 3 & 2.59186 \AA \\
        \hline
        Fe$_2$ & Fe$_3$ & 1 & 2.53864 \AA \\
        \hline
        Fe$_2$ & Fe$_1$ & 3 & 2.73786 \AA \\
        \hline
        Fe$_3$ & Fe$_2$ & 1 & 2.53864 \AA \\
        \hline
        Fe$_3$ & Fe$_2$ & 3 & 2.59186 \AA \\
        \hline
        Fe$_3$ & Fe$_1$ & 3 & 2.35464 \AA \\
        \hline
    \end{tabular}
    \caption{The distances between individual iron sites in \fgt assuming the space group $R\bar3m$.}
    \label{neighbortable}
\end{table}

\subsection{S4 Determination of T$_\text{C}$ using the $J_{ij}$ elements from DFT}
In order to further confirm the validity of our DFT models and calculations, the critical temperature of \fgt has been determined by using both mean-field \cite{Evans.2014} and a Monte Carlo \cite{Skubic2008} methods.

In the mean-field method \cite{Evans.2014} we calculate two separate values for the Curie temperature. The first is an effective 2-dimensional Curie temperature, $T^{2D}_\text{C}$, which we obtain by extracting purely the intra-layer $J_{ij}$ terms which yields $T_\text{C}^\text{2D} = 551 \, \text{K}$. The second is a bulk Curie temperature, $T_\text{C}^\text{3D}$, which additionally contains the inter-layer $J_{ij}$ terms, yielding $T_\text{C}^\text{3D} = 871 \, \text{K}$. While the experimental samples used in the present work are more than $100 \, \text{nm}$ thick, $T_\text{C}^\text{2D}$ might be more appropriate than $T_C^\text{3D}$, because ferromagnetic inter-layer coupling could be mediated by the long-range dipolar interaction. The Monte Carlo method, detailed in Ref. \cite{Skubic2008} yields a Curie temperature of $T^\text{MC}_\text{C} = 500 \, \text{K}$. This implies that the bulk Curie temperature is, as previously suggested, suppressed. 

To conclude, using mean-field and Monte-Carlo methods we have determined three values for the $T_\text{C}$ using the $J_{ij}$ terms determined from \textit{ab initio} methods. We conclude that it is more reasonable to assume the 2D mean-field value describes the system the best because inter-layer coupling could be mediated by the long-range dipolar interaction. The $T_{\text{C}}$ determined from the Monte Carlo method is within a factor of 2 of the measured experimental $T_\text{C}$. As the independent determination of $T_\text{C}$ through the $J_{ij}$ terms is comparable to the experiment, it further justifies their use to determine $K_\delta$ in equations (\ref{KDW_num95}) and (\ref{KDW_num200}).



\subsection{S5 Domain wall widths using the $J_{ij}$ terms from DFT}

\begin{figure}[h!]
    \centering
    \includegraphics[width=\textwidth]{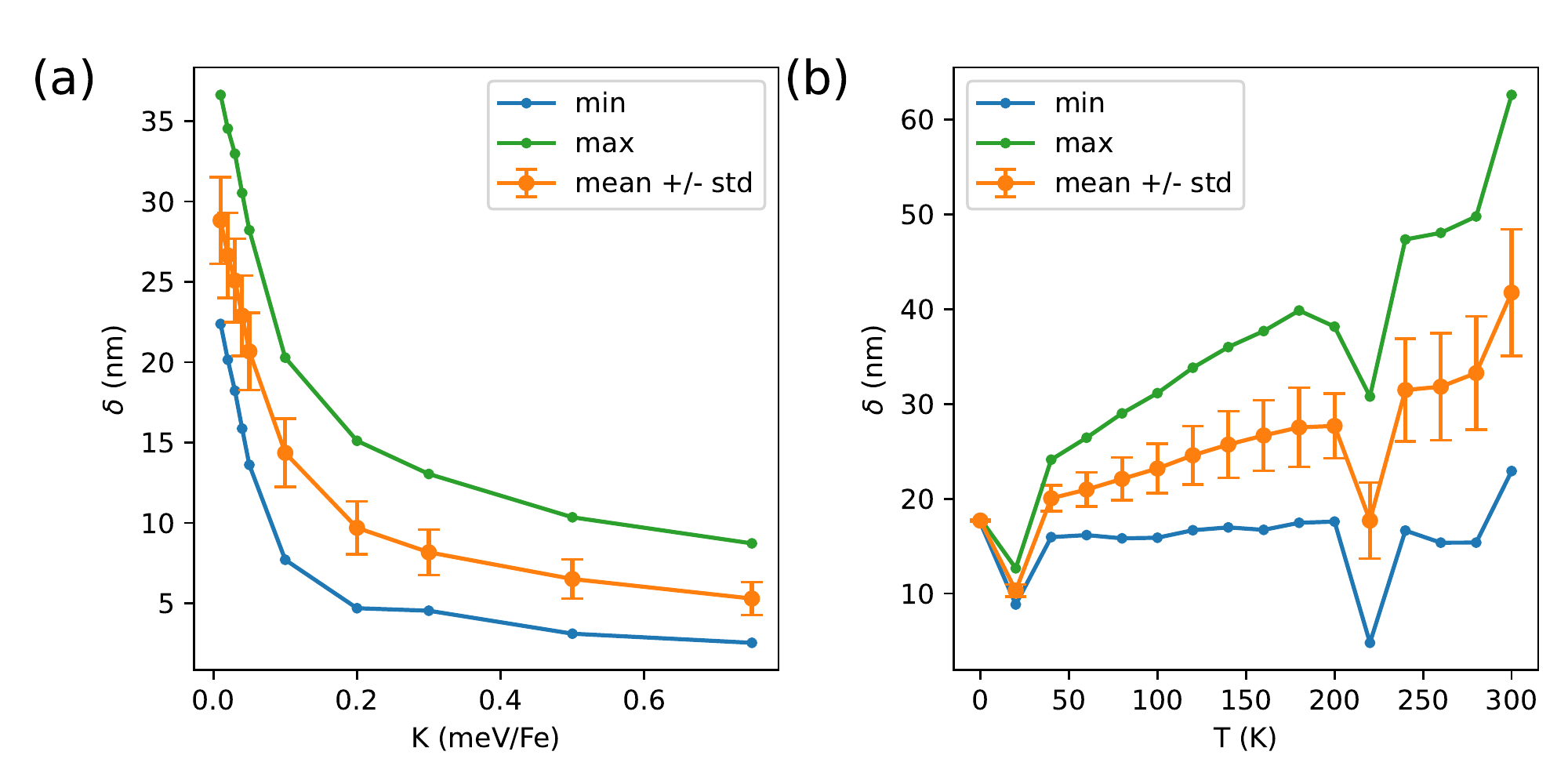} 
    \caption{\textbf{Domain Wall widths using \textit{ab initio} $J_{ij}$'s as a function of anisotropy and temperature}. (a) The domain wall width at $95 \, \text{K}$ as a function of anisotropy constant $K$. (b) The domain wall width with increasing temperature in the simulation. Green and blue lines are indicate the maximal and minimal determined values from the different lines of the Fe atom positions in the unit cell and among the multiple unit cell in the $200\times20\times10$ cell large supercell. The orange line gives the average over the $20\times10$ unit cell cross section parallel to the domain wall.}
    \label{DW_SPIRIT}
\end{figure}

\begin{figure}[h!]
    \centering
    \includegraphics[width=\textwidth]{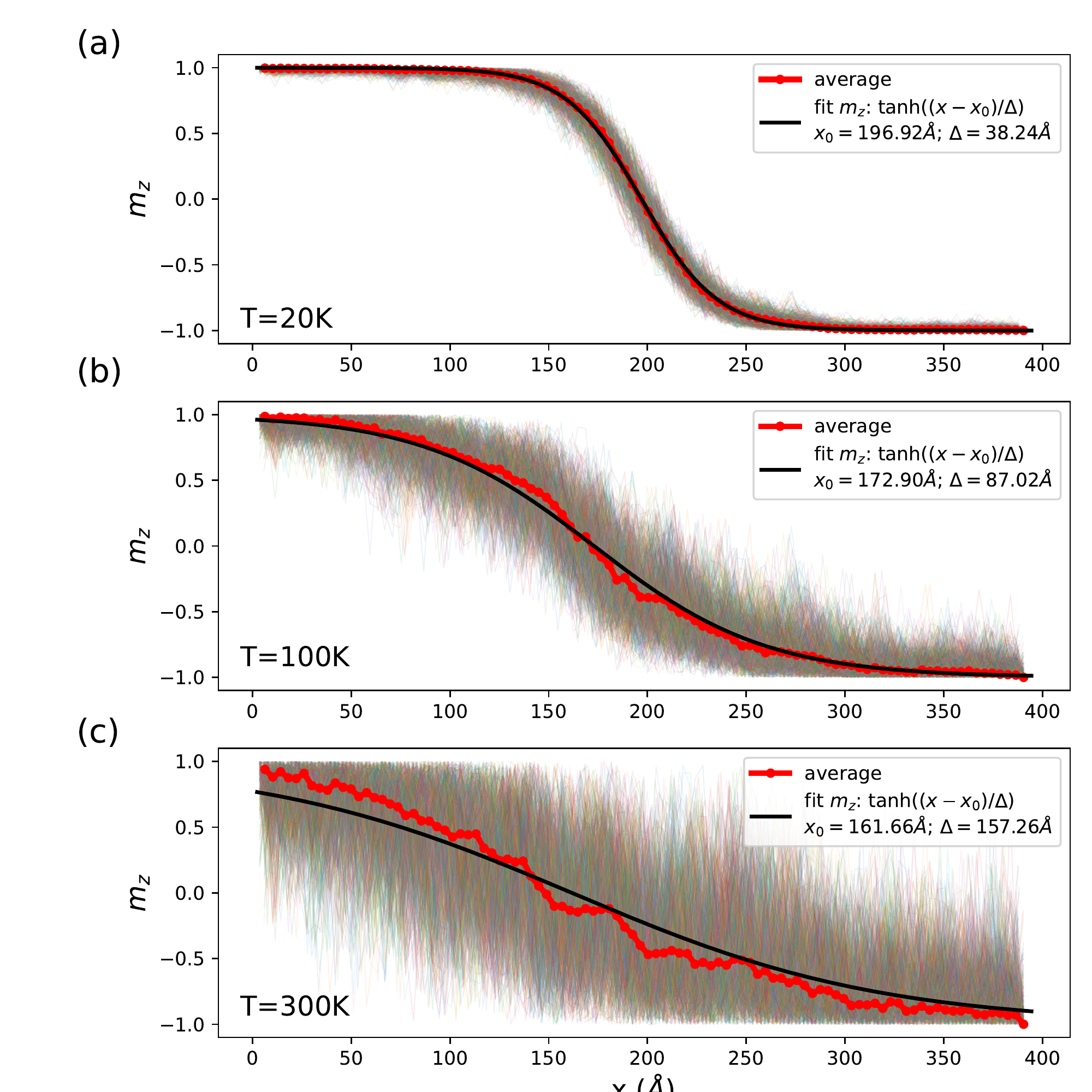}
    \caption{\textbf{Simulated domain Wall profiles}. (a-c) Profiles of the domain wall across all lines of Fe atoms in the $20\times10$ cell cross section of the Spirit supercell for increasing temperature of 20 K, 100 K and 300 K. The faint lines in the background are the inividual line profiles where the temperature fluctations are visible. The thick red lines show the average over the lines and the black line the fitted domain wall profile. The simulation uses the \textit{ab initio} $J_{ij}$'s and the optimized value of the anisotropy of $K=0.04\,\text{eV/Fe}$. For large temperatures the fit starts to deviate from the average line (most visible in c) which is attributed to finite size effects in the simulation.}
    \label{DW_SPIRIT2}
\end{figure}

To further analyse the determination of the exchange parameter $A^\text{DFT}$, we turn to atomistic simulations using the \textit{Spirit} code to use the $J_{ij}$ terms to accurately determine a domain wall width. By simulating a domain wall we can further justify our use of $A^\text{DFT}$ in the determination of the experimental value of $K^{\delta}$. Firstly, we additionally determine $K^\text{DFT}$ directly from the juKKR package giving a value of $K^\text{DFT} = 24 \cdot 10^{-3} \, \text{eV nm}^{-3} \; $. This value is approximately a factor of 30 too large compared to experiment, however this quantity is determined from the bulk when in reality the domain wall is on the surface. This quantity can therefore be considered unknown and hence free to be determined. For our simulations we chose a $100\times20\times10$ supercell consisting of 200,000 spins in the simulation where periodic boundary conditions are chosen in $\vec{b}$ and $\vec{c}$ directions whereas open boundary conditions are used in the $\vec{a}$ direction. The first and last plane of spins in the $\vec{a}$ direction are then fixed in opposite ($\pm z$) directions which enforces a domain wall to form in the center of the simulation box. We solve the Landau-Lifshitz-Gilbert equation with the \textit{Spirit} code \cite{Muller2019} where the Depondt solver is applied for a maximum of 100,000 time steps with a maximal simulation time of $1,000\, \text{ps}$ to relax the magnetic texture. These simulations include the effect of temperature noise which is varied between 0 and 300 K. Fig.~\ref{DW_SPIRIT}(a) is a plot of the domain wall width at $95 \, \text{K}$ for different choices of $K^\text{DFT}$ within the range of experimental values of $K$. For large $K$ t domain wall is shortened significantly compared to the experimental values and is hence not necessary to go beyond 0.7 meV/Fe. As the anisotropy constant decreases we see a rather flat distribution until 0.2 meV/Fe at which point the domain wall diverges for vanishing $K$. Choosing a particular value for $K=0.04\,\text{meV/Fe}$  to match with the experimental value, Fig.~\ref{DW_SPIRIT}(b) shows how the domain wall evolves as a function of temperature. We observe a steadily increasing value for the domain wall width with temperature, crossing $\delta=23.2\,\text{nm}$ at $100 \text{K}$ and $\delta=27.7\,\text{nm}$ at $200 \text{K}$. This is in good agreement with experimentally determined domain walls when using the experimental anisotropy instead of the value determined by DFT which is more specific to the bulk. Three exemplary domain wall fit at temperatures of 20 K, 100 K and 300 K are shown in Fig.~\ref{DW_SPIRIT2}(a-c).

\end{document}